\documentclass{ws-procs11x85}
\usepackage{ws-procs-thm}
\usepackage{hyperref}
\usepackage{graphicx}
\let\captionbox\undefined
\usepackage{subcaption}
\graphicspath{ {./images/} }
\usepackage{multicol}
\usepackage{mathrsfs}
\usepackage{amsmath}
\usepackage{amssymb}
\usepackage{esint}
\usepackage{algorithm}
\usepackage{algorithmicx}
\usepackage[noend]{algpseudocode}

\newcommand\cP{\mathcal{P}}

\newcommand\cX{\mathcal{X}}
\makeatletter
\newcommand*{\algolabel}[2]{%
  \hypertarget{#1}{}%
  \NR@gettitle{#1}%
  \label{#2}%
}
\makeatother
\newcommand\ch[1]{\mathbf{ch}\left(#1\right)}
\newcommand\pr[1]{\nu_{#1}}
\newcommand\V[1]{\mathbf{#1}}
\makeatletter
\def\@fnsymbol#1{\ensuremath{\ifcase#1\or 1 \or 2 \else\@ctrerr\fi}}
\makeatother
\newcommand\DP[0]{\mathbf{DP}}
\catcode`\%=12
\newcommand\pcnt{
\catcode`\%=14
\setlength{\belowcaptionskip}{-10pt}

\begin{document}

\title{Efficient Reconstruction of Stochastic Pedigrees:\\Some Steps From Theory to Practice}
\author{Elchanan Mossel and David Vulakh$^*$}
\address{Department of Mathematics, Massachusetts Institute of Technology, \\
    77 Massachusetts Ave, Cambridge, MA 02139, USA \\
    $^*$E-mail: dvulakh@mit.edu
}
\copyrightinfo{Preprint of an article published in Pacific Symposium
  on Biocomputing \copyright\ 2022 World Scientific
Publishing Co., Singapore, \url{http://psb.stanford.edu/}}

\begin{abstract}
In an extant population, how much information do extant individuals
provide on the pedigree of their ancestors? Recent work by Kim,
Mossel, Ramnarayan and Turner (2020) studied this question under a
number of simplifying assumptions, including random mating, fixed
length inheritance blocks and sufficiently large founding population.
They showed that under these conditions if the average number of
offspring is a sufficiently large constant, then it is possible to
recover a large fraction of the pedigree structure and genetic content
by an algorithm they named REC-GEN.

We are interested in studying the performance of REC-GEN on simulated
data generated according to the model.  As a first step, we improve
the running time of the algorithm.  However, we observe that even the
faster version of the algorithm does not do well in any simulations in
recovering the pedigree beyond 2 generations.  We claim that this is
due to the inbreeding present in any setting where the algorithm can
be run, even on simulated data.  To support the claim we show that a
main step of the algorithm, called ancestral reconstruction, performs
accurately in an idealized setting with no inbreeding but performs poorly in random mating populations. 

To overcome the poor behavior of REC-GEN we introduce a Belief-Propagation based heuristic that accounts for the inbreeding and performs much better in our simulations.  
\end{abstract}

\section{Introduction}\label{sec:intro}

We follow up on a recent
work by Kim et al~\cite{kim2020efficient}., the main motivation of
which
is to understand how much kinship
information can be learned from DNA.
More concretely, Kim et al. study the inference problem of recovering
ancestral kinship relationships of a population of \textit{extant}
(present-day) individuals using only their genetic data for a
mathematical generative model of pedigrees and DNA sequences on them
based on the combinatorial framework of Steel and
Hein\cite{steel2006reconstructing} and Thatte and
Steel\cite{thatte2008reconstructing}, who also proved a rigorous
statement about recovery of idealized pedigree models. The goal is to
use this extant genetic data to recover the \textit{pedigree} of the
extant population under this model.


To study this question, Ref.~\citen{kim2020efficient} introduces an
idealized model for generating pedigree data. The population model
they use is a standard random mating model, but the genetic
inheritance model assumes that inheritance blocks are of fixed length.
This removes the additional difficulty of ``phasing" which allows for
a rigorous analysis.


The main contribution of Ref.~\citen{kim2020efficient} is to show that
under certain conditions, the algorithm proposed in the paper, named
{\sc Rec-Gen}, approximately recovers the true, unknown pedigree as
well as its genetic content. There is a huge body of work on pedigree
reconstruction, see e.g.
Ref.~\citen{thompson00,kirkpatrick2011pedigree,IPED,thompson2013identity,IPED2,shem2014historical,huisman2017pedigree,wang2019pedigree}.
In contrast to Ref.~\citen{kim2020efficient}, most of this work does not provide theoretical guarantees. In this
paper we take the theoretical analysis in
Ref.~\citen{kim2020efficient} and study to what extent it can be
applied in more realistic settings.

There is a tension between different aspects of the assumptions in
Ref.~\citen{kim2020efficient}. 
On one hand, they require a very big pedigree to avoid inbreeding. On
the other, the algorithm Rec-Gen has cubic running time. While in the
limit as the pedigree size goes to infinity, this tension disappears,
we find that applying the algorithm on simulated data results either
in poor accuracy or an infeasible running time. 

Our main contributions in this paper are:
\begin{itemize}
\item 
We improve the algorithm runtime to essentially quadratic for
model-generated data.
\item 
We then observe that even the faster version of the algorithm does not do well in any simulations in recovering the pedigree beyond 2 generations. 
\item 
We claim that this is due to the inbreeding present in any setting where the algorithm can be run, even on simulated data. 
\item 
To support the claim we show that a main step of the algorithm, called ancestral reconstruction, performs accurately in a setting with no inbreeding but performs poorly in random mating populations. 
\item 
Finally, to overcome the poor behavior of REC-GEN we introduce a Belief-Propagation based heuristic that accounts for the inbreeding and performs much better in our simulations. 
\end{itemize}

\section{Model Description}\label{sec:model}

We model populations as in Ref.~\citen{kim2020efficient}. Here, we briefly
restate the definition of a \emph{coupled pedigree}, the structure
manipulated by the \textsc{Rec-Gen} algorithm  and introduce notation relevant to the
description of our modified algorithm.

A $(N,B,T,\xi)$-\emph{uncoupled pedigree} $\mathcal{U}$ is a directed acyclic
graph $(V, E)$ in which vertices $v \in V$ represent individuals and
edges $e = (u, v) \in E$ represent the relationship that $u$ is a
\emph{parent} of $v$. The set of vertices $V$ can be partitioned into
$T + 1$ subsets $V_0,\dots,V_T$ so that each $v \in V_i$,
$0 \le i < T$ has exactly two in-edges, both of which are from
vertices in $V_{i + 1}$. The sets $V_i$ represent \emph{generations},
where $V_0$ is the \emph{extant population} and $V_T$ is the
\emph{founding population}. The size of the founding population
$|V_T|$ equals $N$. The vertices $V$ also satisfy \emph{monogamy} ---
within each generation $V_i$, $i > 0$, the vertices $V_i$ can be
partitioned into pairs $(v_1, v_2)$ such that if $u$ is a child of
$v_1$ if and only if $u$ is a child of $v_2$; such pairs are called
\emph{couples}. The number of children of each couple is randomly
drawn from the distribution $\xi$.

Each vertex $v$ has associated genetic information in the form of $B$
\emph{blocks}, each of which contains a symbol sampled from some
alphabet $\Sigma$. The symbol at block $b$ of the genome of vertex
$v$ is denoted $s_v(b)$.

The $(N,B,T,\xi)$-\emph{coupled pedigree} $\mathcal{P}$ induced by an
uncoupled pedigree $\mathcal{U}$ is formed by merging each couple into
a single vertex --- the resulting vertices are called \emph{coupled
nodes}. Now each $v$ in a generation other than the extant represents
a pair of individuals, and an edge from a coupled node $u$ to a
coupled node $v$ represents that $u$ is the parent of one of the
individuals in couple $u$. All vertices in $\mathcal{P}$ have
in-degree two, except vertices in the extant population, which remain
uncoupled and have in-degree 1. The genetic information $s_v(b)$ of a
coupled node $v$ is the set of all symbols that are in
block $b$ for some individual in the couple represented by $v$.

When we say that some graph is a pedigree in this paper without
specifying whether the pedigree is coupled or uncoupled, we are
referencing coupled pedigrees.

\section{{\sc Rec-Gen}}\label{sec:kim2020efficient}

The \textsc{Rec-Gen} reconstruction algorithm presented in
Ref.~\citen{kim2020efficient} proceeds in three main phases.  In
each generation, siblinghood detection reconstructs relationships in the current generation,
outputting a siblinghood hypergraph in which triple $u, v, w$ forms a
hyperedge if they are likely to be siblings. Parent construction
processes maximal cliques in the outputted hypergraph, populating the
parent generation. Symbol collection reconstructs the genetic
information of the parent generation.
\subsection{Runtime Analysis}\label{sec:runtime}

Na\"ive implementations of siblinghood detection and symbol collection
both run in $\Omega(BN_0^3)$ time.
The siblinghood test counts the number of shared
blocks in all triples, which can require $\Omega(BN_0^3)$ in the worst
case.

To na\"ively find a triple of extant vertices sharing a gene in block
$b$ with $u$ as their joint-LCA for some $u$ in generation $t$ of the
pedigree for symbol collection it may be
necessary to inspect all triples of extant descendants of $u$ in each
block $b$, which is also $\Omega(BN_0^3)$.

We wish to improve both of these processes to $O(BN_0^2)$, as
described in Sections~\ref{sec:fast-sib} and~\ref{sec:fast-symb}.

\subsection{Faster Siblinghood Detection}\label{sec:fast-sib}

The greatest bottleneck in the runtime of \textsc{Rec-Gen} is the
siblinghood detection step, which for each generation $t$ is cubic in
the size of that generation $N_t$. To reduce the runtime from
$O\left( N_t^3B \right)$ to $O\left( N_t^2B\right)$, we begin by
processing all pairs of vertices, marking pairs that share some
threshold $\theta=0.4$ of their blocks as \emph{sibling candidates}.
We then only consider triples of vertices formed from sibling
candidates when generating the siblinghood hypergraph. Pseudocode of
the alternate algorithm \nameref{algo:fast-test-siblings} follows:

\begin{algorithm}[t]
\caption{
    Perform statistical tests to detect siblinghood
}
\begin{algorithmic}[1]
\algolabel{\textsc{Fast-Test-Siblinghood}}{algo:fast-test-siblings}
\Procedure{Fast-Test-Siblinghood}{depth $(k-1)$ pedigree $\hat{\cP}$}
    \State $C \gets \varnothing$
    \State $V \gets$ vertices of $\hat{\cP}$ at level $k-1$
    \ForAll{distinct pairs $\{u,v\} \in 2^V$}
        \If{
            $\geq 0.4|B|$ blocks $b$ such that
            $\hat{s}_{u}(b) \cap \hat{s}_{v}(b) \neq \varnothing$
        }
            \State $C \gets C \cup \{u,v\}$
        \EndIf
    \EndFor
    \State $E \gets \varnothing$
    \ForAll{pairs $\{u,v\} \in C$}
        \ForAll{
            $w \in V$ at level $k-1$ such that
            $w \neq u \land w \neq v$
        }
            \If{
                $\geq 0.21|B|$ blocks $b$ such that
                $
                    \hat{s}_{u}(b) \cap
                    \hat{s}_{v}(b) \cap
                    \hat{s}_{w}(b) \neq \varnothing
                $
            }
                \State $E \gets E \cup \{u,v,w\}$
            \EndIf
        \EndFor
    \EndFor
    \State $\hat G \gets (V, E)$
    \State \Return $\hat{G}$
\EndProcedure
\end{algorithmic}
\end{algorithm}


\subsection{Faster Symbol Collection}\label{sec:fast-symb}

To decrease the complexity of executing the \textsc{Rec-Gen}
symbol-collection phase on $v$, we avoid explicitly searching for
extant triples that have $v$ as their joint-\textsf{LCA}. Instead, we
make the simplifying assumption that any three extant vertices $x,y,z$
descended from distinct children of $v$ have $v$ as a
joint-\textsf{LCA}. Now, we can use the following modified algorithm
to achieve an effect equivalent to the original
symbol collection of Ref.~\citen{kim2020efficient}:
\begin{itemize}
    \item Let $\hat{G}_u(b)$ for a child $u$ of $v$ and block $b$ be
        the set of genes $g$ such that there exists an extant
        descendant $x$ of $u$ such that $\hat{s}_u(b) = \{g\}$.
    \item Compute $\hat{G}_u(b)$ for all children $u$ of $v$.
    \item Let $\hat{s}_v(b)$ be the two genes that are present in the
        greatest number of computed sets $\hat{G}_u(b)$.
\end{itemize}
Pseudocode of this modified process can be seen in Algorithm
\nameref{algo:fast-symbols}.

Ref.~\citen{kim2020efficient} prove that, conditioned on the nonoccurence of
undesirable inbreeding events, the existence of a joint-\textsf{LCA}
$v$ for three nodes $x,y,z$ entails that $v$ is their unique
\textsf{LCA}. Therefore, if most extant nodes have a
joint-\textsf{LCA}, then the algorithm described above is equivalent
to the initial description of symbol-collection.
Empirically, very few ($<1\%$ of) extant triples in simulated
pedigrees are descended from unique children of a vertex that is not
their joint-\textsf{LCA}.

Generating $\hat{G}_u$ requires time that is linear in the number of
nodes in the descendants pedigree of $u$. Since $\alpha > 2$, this is
on expectation bounded above by a linear function of the number of
extant descendants of $u$. Each extant individual $v$ has at most
$2^t$ ancestors in generation $t$. Therefore, the sum of the number of
extant descendants of $u$ over all $u$ in generation $t$ is at most
$2^tN_0$, where $N_0$ is the size of the extant population, so that
the runtime of invoking Algorithm~\nameref{algo:fast-symbols} for all
$u$ at generation $t$ is $O(B \cdot (2^tN_0 + |G|))$. Since
$\alpha > 2$,
$2^t \subseteq O(\alpha^t) \subseteq O(\mathbb{E}[N_t]/N_T)$, so that
the total runtime of Algorithm~\nameref{algo:fast-symbols} is
$O(B \cdot \mathbb{E}[N_t]N_0/N_T) \subseteq O(BN_0^2)$.

\begin{algorithm}[t]
\caption{
    Empirically reconstruct the symbols of top-level node $v$ in
    $\cP$.
}
\begin{algorithmic}[1]
\algolabel{\textsc{Fast Collect-Symbols}}{algo:fast-symbols}
\Procedure{Fast Collect-Symbols}{$v, \hat{\cP}$}
    \ForAll{blocks $b \in [B]$}
        \State $c_g \gets 0 \;\forall\; g$
        \ForAll{children $u$ of $v$}
            \State $\hat{G}_u(b) \gets \varnothing$
            \ForAll{extant $x$ descended from $u$}
                \State $\hat{G}_u(b) \gets \hat{G}_u(b) \cup \hat{s}_x(b)$
            \EndFor
            \ForAll{$g \in \hat{G}_u(b)$}
                \State $c_g \gets c_g + 1$
            \EndFor
        \EndFor
        \State $\sigma_1 \gets g\text{ with highest }c_g$
        \State $\sigma_2 \gets g\text{ with second-highest }c_g$
        \State Record the symbols $\sigma_1, \sigma_2$ for block $b$ in $v$.
    \EndFor
\EndProcedure
\end{algorithmic}
\end{algorithm}

\section{Simulations}\label{sec:simulations}

We assess the empirical accuracy of \textsc{Rec-Gen} and other
algorithms presented later in this work by running them on simulated
pedigrees. We generate pedigrees satisfying the stochastic model, as
described in section \ref{sec:gen-ped}. The extant populations of the
pedigrees can be used as input for our implementations of the
reconstructive algorithms, and a grader program evaluates the accuracy
of the result as described in section~\ref{sec:grader-prog}.

\subsection{Generating Pedigrees}\label{sec:gen-ped}

For a given $\alpha$, our pedigree generator program creates
$(N,B,T,\xi)$-coupled pedigrees according to the breeding and
inheritance behaviors described in Section~\ref{sec:model}, where
$\xi$ is either Poisson-distributed with parameter $\alpha$ or a
constant distribution, $\xi = \alpha$.

\subsection{Assessing Reconstruction Accuracy}\label{sec:grader-prog}

Our grader program takes as input a parameter $\alpha \in [0, 1)$ and
two pedigrees with identical extant populations and the same numbers
of generations | an original pedigree $\cP$ and its reconstruction
$\cP'$. It outputs a partial mapping between the coupled nodes of
$\cP$ and $\cP'$, where a coupled node $v \in \cP$ of the original
pedigree is mapped to a coupled node $v' \in \cP'$ of the
reconstructed pedigree only if $v'$ is an \emph{$\alpha$-successful}
reconstruction of $v$. $\alpha$-successful reconstructions are
defined recursively in the following manner:
\begin{itemize}
    \item In generation 0 (the extant population) a vertex
        $v' \in \cP'$ is an $\alpha$-successful reconstruction of
        $v \in \cP$ if and only if $v$ and $v'$ are the same coupled
        node.
    \item In generation $t > 0$, let $c(v, v')$ for
        $v \in \cP, v' \in \cP'$ denote the number of pairs $u$ and
        $u'$ from generation $t - 1$ of $\cP$ and $\cP'$,
        respectively, for which $u$ is a child of $v$, $u'$ is a
        child of $v'$, and $u'$ is an $\alpha$-successful
        reconstruction of $u$. Also, let $f$ be the number of
        children of $v$ and $f'$ be the number of children of $v'$.
        Then $v'$ is an $\alpha$-successful reconstruction of $v$ if
        and only if $c(v, v') > \alpha f$ and $c(v, v') > \alpha f'$.
\end{itemize}
In the case that multiple vertices $v' \in \cP'$ are an
$\alpha$-successful reconstruction of some $v \in \cP$, the grader
program maps $v$ to the one that maximizes $c(v, v')$. If the program
maps some $v'$ to $v$, we consider $v$ successfully reconstructed.

The grader also outputs the
following statistics for each generation $t$:
\begin{itemize}
    \item The number and percent of successfully reconstructed
        vertices
    \item The number and percent of successfully reconstructed edges
        (these are the sum of $c(v, v')$ over reconstructed $v$ and
        the ratio of that sum to the sum of $f$ over all $v$ in $t$)
    \item The number of reconstructed blocks, where a block
        $g$ in position $b$ of $v$ is considered reconstructed if
        $v'$ also has $g$ in position $b$\footnote{In case that $v$
        has two identical genes in some position, they are both
        considered reconstructed only if $v'$ also has two copies of
        that gene; otherwise, only one is considered reconstructed.
        Note, however, that this should not happen regularly, as it
        is an indication of inbreeding.}, as well as the percent of
        blocks reconstructed out of all blocks in generation
        $t$ and out of blocks belonging to reconstructed nodes in
        generation $t$.
\end{itemize}

When using the grader to study the behavior of the reconstruction of
symbols, we typically apply a generous $\alpha = 0.5$ threshold, so
as not to exclude information about weakly reconstructed vertices.
For the accuracy metrics presented throughout this paper, we usually
use $\alpha = 0.75$ or $\alpha = 0.99$.

\section{Simulation Results for {\sc Rec-Gen}}\label{sec:kim2020efficient-sim}

\subsection{Results for\texorpdfstring{
    $\boldsymbol{T = 3}$}{T = 3}
}\label{sec:kim2020efficient-sim-t3}

\begin{figure}
    \centering
    \begin{subfigure}{.475\textwidth}
        \centering
        \includegraphics[width=\textwidth]{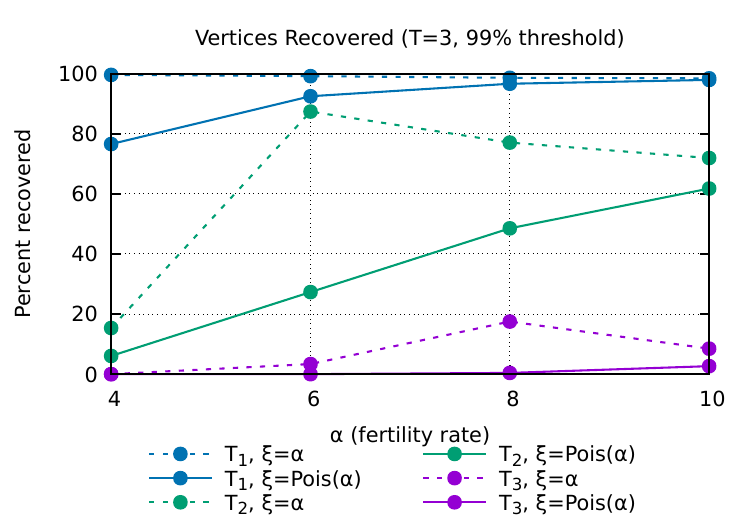}
    \end{subfigure}
    \begin{subfigure}{.475\textwidth}
        \centering
        \includegraphics[width=\textwidth]{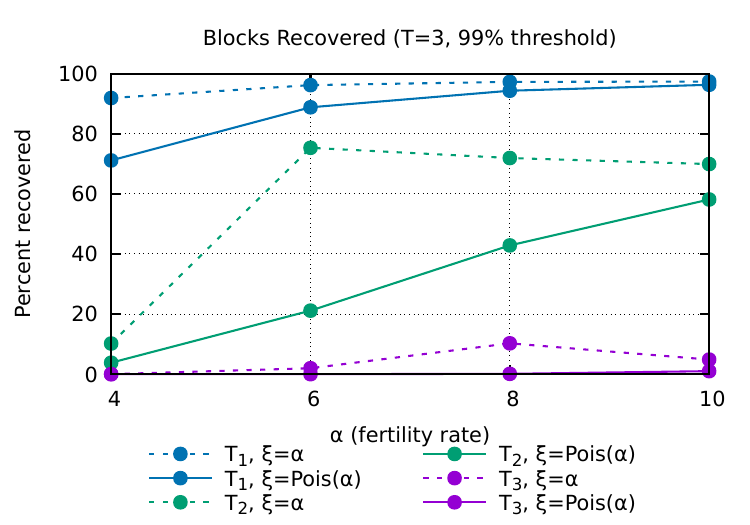}
    \end{subfigure}
    \caption {
        Average percent vertices and blocks 0.99-successfully reconstructed in
        each generation
    }\label{fig:vert-t3-a99}
\end{figure}

\begin{figure}
    \centering
    \begin{subfigure}{.475\textwidth}
        \centering
        \includegraphics[width=\textwidth]{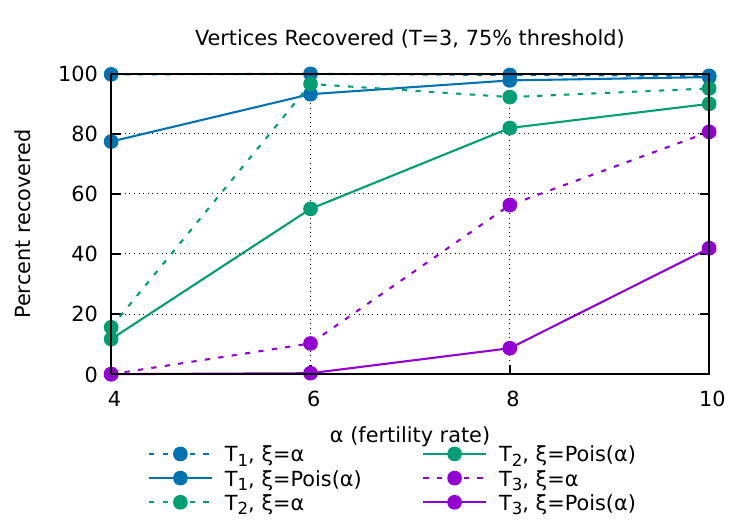}
    \end{subfigure}
    \begin{subfigure}{.475\textwidth}
        \centering
        \includegraphics[width=\textwidth]{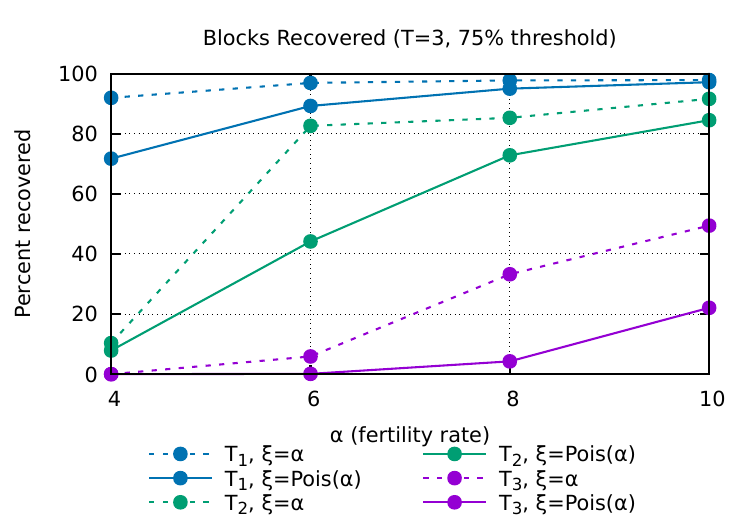}
    \end{subfigure}
    \caption {
        Average percent vertices and blocks 0.75-successfully reconstructed in
        each generation
    }\label{fig:vert-t3-a75}
\end{figure}

Experiments using simulated data as described in Section
\ref{sec:gen-ped} indicate that, even in pedigrees with relatively
small founding populations ($N=50$) and fertility rates ($\alpha=6$),
\textsc{Rec-Gen} reliably reconstructs two generations above the
extant (the `parent' and `grandparent' generations) in pedigrees with
$T=3$. However, performance at the third generation declines sharply,
and in individual simulations with $T=4$ (not included in the batched
results in this section; see Section \ref{sec:kim2020efficient-sim-t4}),
\textsc{Rec-Gen} fails to recover even a single vertex of the founding
population. Figures~\ref{fig:vert-t3-a99}
and~\ref{fig:vert-t3-a75} graph the average vertices and blocks
reconstructed over $\alpha$ for three generation pedigrees with $N=50$
and $B=5000$ for two values of the reconstruction accuracy threshold:
$0.75$ and $0.99$.

As one would expect, reconstruction accuracy generally improves as
$\alpha$ increases (an exception is for the high accuracy threshold
$0.99$ in the case of constant fertilities --- when there is a larger
number of children, even an algorithm that reconstructs each with
higher probability may reconstruct all of them with lower
probability). Additionally, \textsc{Rec-Gen} performs better for the
case of constant fertilities than for the case of Poisson-distributed
fertilities. Since \textsc{Rec-Gen} performs poorly for vertices with
low fertility (and is incapable of reconstructing vertices with
fertility less than 3), we attribute the relatively poor performance
of \textsc{Rec-Gen} for the Poisson case as compared to the
deterministic case to the incidence of low-fertility nodes.

\subsection{Decline at\texorpdfstring{
    $\boldsymbol{T = 4}$}{T = 4}
}\label{sec:kim2020efficient-sim-t4}

\begin{figure}
    \centering
    \begin{subfigure}{.45\textwidth}
        \centering
        \includegraphics[width=\textwidth]{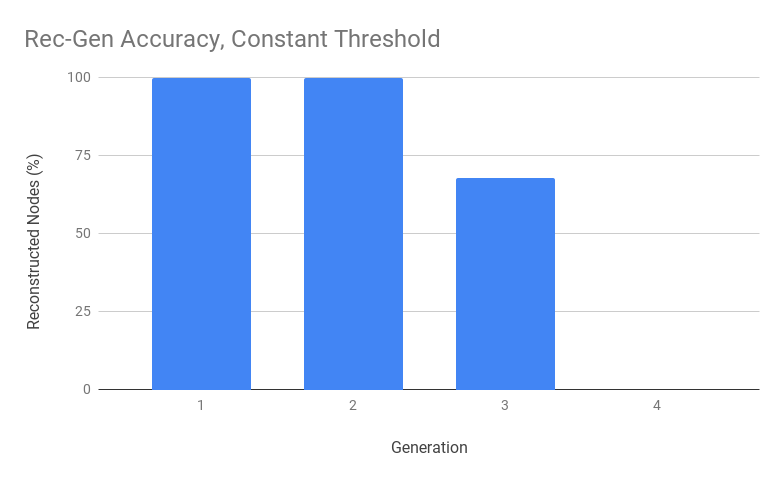}
        \caption{Constant threshold}
    \end{subfigure}
    \begin{subfigure}{.45\textwidth}
        \centering
        \includegraphics[width=\textwidth]{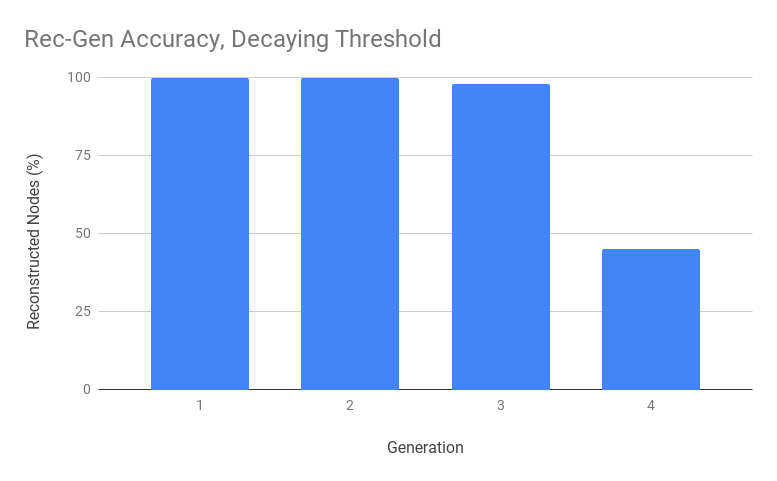}
        \caption{Adjusted threshold}
    \end{subfigure}
    \caption{
        Vertices 0.5-reconstructed by \textsc{Rec-Gen} for a $T=4$
        pedigree, with both the default $21\%$ siblinghood threshold
        (a) and a manually optimized siblinghood threshold (b). Note
        that 0 nodes are reconstructed in generation 4 in (a).
    }
    \label{fig:node-decay}
\end{figure}

\begin{figure}
    \centering
    \begin{subfigure}{.24\textwidth}
        \centering
        \includegraphics[width=\textwidth]{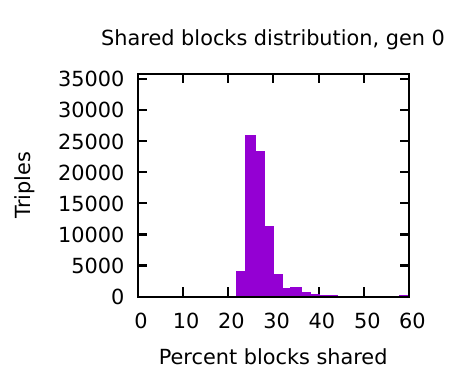}
        \caption{Extant}
    \end{subfigure}
    \begin{subfigure}{.24\textwidth}
        \centering
        \includegraphics[width=\textwidth]{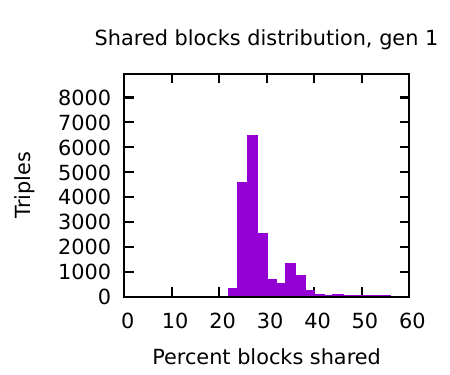}
        \caption{Parents}
    \end{subfigure}
    \begin{subfigure}{.24\textwidth}
        \centering
        \includegraphics[width=\textwidth]{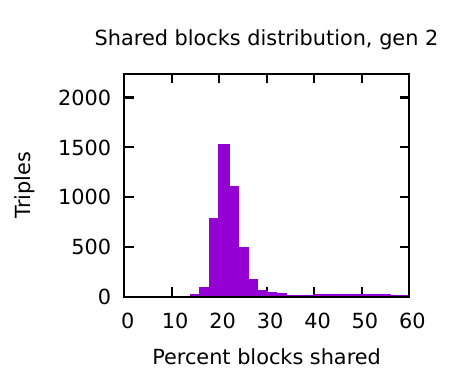}
        \caption{Grandparents}
    \end{subfigure}
    \begin{subfigure}{.24\textwidth}
        \centering
        \includegraphics[width=\textwidth]{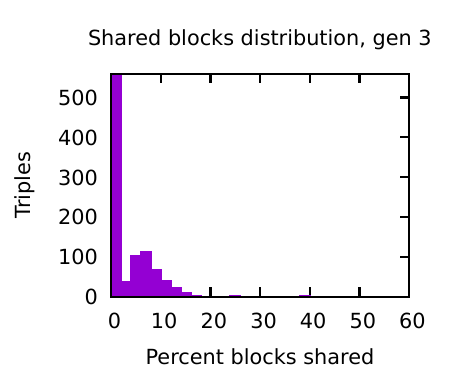}
        \caption{Founder's children}
    \end{subfigure}
    \caption {
        Distribution of percent reconstructed blocks shared in all
        triples for a $T=4$ pedigree.
    }\label{fig:block-decay}
\end{figure}

As demonstrated in Figure~\ref{fig:node-decay} (a), \textsc{Rec-Gen}
appears to encounter major difficulties by the fourth generation,
failing to recover even a single founding node in our simulations.
This failure seems to be precipitated by a rapid decline in accuracy
of reconstructed blocks, as shown in Figure~\ref{fig:block-decay}.
Recall that symbol collection requires that
triples share at least $21\%$ of their blocks to be identified as
siblings. In generations 0 and 1, the distribution of shared
reconstructed blocks for sibling triples lies entirely above the
$21\%$ threshold. By generation 2, it shifts slightly to the left so
that some siblings are not recognized (and, as a result, not all of
the generation 3 is reconstructed). In generation 3, there are two
clusters in the distribution of shared triples: one at 0\% and one
around 10\%. The cluster at 0\% is the result of the members of
generation 3 who were not reconstructed at all; the rest of the
distribution consists of the remaining triples, which still share
distinctly more blocks than non-sibling triples, but fewer than 21\%.

When we manually set the siblinghood threshold to decay with each
generation to match the accumulation of errors, we can extend the
number of generations for which \textsc{Rec-Gen}
accurately reconstructs the topology. Figure~\ref{fig:node-decay} (b)
demonstrates the improvements when using the siblinghood
thresholds 21\%, 21\%, 17\%, 4\% for generations 0, 1, 2, and 3
respectively.

This experiment implies that the step that introduces the most error
into \textsc{Rec-Gen} is the symbol-collection step. In reality, we
cannot easily manually adjust the siblinghood threshold, because the
optimal threshold varies from pedigree to pedigree and may be
difficult to determine without knowledge of the true pedigree
topology. We can further assume that these errors are largely the
result of failure of the combinatorial \textsc{Rec-Gen} algorithm to
correctly handle inbreeding. We confirm this assumption by running
\textsc{Rec-Gen} on a large pedigree constructed as though it were a
section sampled from an infinitely wide pedigree --- indeed,
\textsc{Rec-Gen} has almost perfect accuracy in this case, as expected
(the only errors were the result of blocks that were not passed down
to any descendants, which can happen with frequency $1/2^\alpha$). We
therefore wish to improve the robustness of the symbol-collection step
against inbreeding.

\section{Belief Propagation}\label{sec:bp}

To improve the empirical accuracy of the symbol collection step, we
replace the original combinatorial
symbol collection algorithm with a single pass of a
Belief-Propagation (BP) algorithm for recovering the genetic
information of pedigrees. BP is a message-passing algorithm for
inference that is most successful in locally tree-like models.
Mezard and Montanari~\cite{mezard2009information} give the BP equations in the following setting:
\begin{itemize}
    \item $\V x$ is a tuple of $N$ variables
        $\left( x_1, \dots, x_N \right)$ assuming values from the
        finite alphabet $\cX$.
    \item There are $M$ constraints in the form of the marginals
        $\psi_1, \dots, \psi_M$ governing the distribution of values
        assumed by $\V x$, so that the probability distribution of $x$
        satisfies
        \[
            p(\V x) \cong \prod_{a = 1}^M \psi_a\left( 
                \V x_{\partial a}
            \right)
        \]
        where
        $
            \V x_{\partial a} = \left\{
                x_i : i \in \partial a
            \right\}
        $
        and $\partial a \subseteq [N]$ is the set of variable indices
        constrained by $\psi_a$ (here, the notation $x \cong y$
        denotes that the two functions $x,y : \cX \to \mathbb{R}$ are
        equal down to a constant factor).
\end{itemize}
In this context, the relationships between variables can be modelled
by a bipartite graph in which each vertex representing a variable
$x_i$ has an edge to each `factor vertex' representing a constraint
$\psi_a : i \in \partial a$; this graph is called the
\emph{factor graph}. The BP equations that permit approximation of the
marginal distribution of each variable govern `messages' sent over the
edges of the factor graph at each time step $t + 1$:
\begin{itemize}
    \item Message from the $j$th variable to the $a$th factor:
        \[
            \nu_{j \to a}^{(t + 1)}\left( x_j \right) \cong
                \prod_{b \in \delta j \setminus a}
                \widehat\nu_{b \to j}^{(t)}\left( x_j \right)
        \]
    \item Message from the $a$th constraint to the $j$th variable:
        \[
            \widehat\nu_{a \to j}^{(t)}(x_j) \cong
                \sum_{\V x_{\partial a \setminus j}} \psi_a\left(
                    \V x_{\partial a}
                \right) \prod_{k \in \partial a \setminus j}
                    \nu_{k \to a}^{(t)}\left( x_k \right)
        \]
\end{itemize}
The estimate for the marginal distribution of variable $i$ at time
$t$ is
\[
    \nu_i^{(t)}\left( x_i \right) \cong
        \prod_{a \in \partial i} \widehat\nu_{a \to i}^{(t - 1)}
        \left(
            x_i
        \right)
\]
If the factor graph is a tree, then BP is known to be exact | that is,
the values $\nu_i$ converge, and they converge precisely to the true
marginals of the variables. Moreover, the exact marginals can be
computed with BP in linear time in the tree case, as $\nu_i$ assume
the values of the marginals of $x_i$ after two passes through the
tree, as described in Ref.~\citen{mezard2009information}.

For our modified symbol-collection step, we effectively complete one
BP sweep (half of the tree algorithm) independently for each position
in the genome. Let $G$ be the set of all genes, $\ch{v}$ be the tuple
of children of vertex $v$, $0 < \varepsilon < 1$ be some constant that
represents the probability of an error in the topology of the
reconstruction, $\V g_v$ be the variable the value of which is the
pair of genes in a given block of vertex $v$, and
$\pr{v}$ be a function from unordered pairs from $G$ to the unit
interval, the BP estimate of the marginal distribution of $\V g_v$.

For each extant vertex $v$ with gene $g$, we introduce a constraint
\[
    \psi \cong \mathbb{I}_{\V g_v = (g, g)}
\]
For each nonextant vertex $v$, we introduce a constraint indicating
that a child of $v$ is an anomaly in the topology (shares no genes
with $v$) with probability $\varepsilon$:
\[
    \psi \cong
        \varepsilon^{\scalebox{0.75}{$
            |\{u \in \ch{v} :
            \V g_v \cap \V g_u = \varnothing\}|
        $}}
\]
Then the computed values of $\pr{v}$ are as follows. For extant
couples $v$ with gene $g$, we have
\[
    \pr{v}(g_1, g_2) \cong \mathbb{I}_{g_1 = g_2 = g}
\]
And for nonextant couples $v$
\[
    \pr{v}(g_1, g_2) \cong
        \sum_{\V{g}\in (G^2)^{|\ch{v}|}}
        \varepsilon^{\scalebox{0.75}{$
            |\{i \in [1, |\ch{v}|] :
            \{g_1, g_2\} \cap \V{g}_i = \varnothing\}|
        $}}
        \prod_{i \in [1,|\ch{v}|]} \pr{\ch{v}_i}(\V{g}_u)
\]
We record the gene pair with the highest probability according to
$\pr v$ as the genes reconstructed for couple $v$.

Computing $\pr{v}$ directly would be computationally inefficient |
worse than $O\left( |G|^{2\alpha} \right)$ on expectation, as
$|\ch{v}|$ is Poisson-distributed with parameter $\alpha$. We can
substantially improve this runtime by computing the probability
distribution by summing over the number of children indicating
topology errors, rather than over all possible assignments of genes.
To do this, we construct a DP table $\DP(g_1, g_2)_{i,j}$ that stores,
for the first $i$ children, the probability that $j$ of them indicate
topology errors. The recursive definition follows:
\[
    \DP(g_1, g_2)_{i,j} = \left\{\;\begin{aligned}
        & \mathbb{I}_{j = 0} & i = 0 \\
        & \DP(g_1, g_2)_{i-1,j-1}\sum_{(h_1, h_2) \in G^2}
            \mathbb{I}_{\{h_1, h_2\} \cap \{g_1, g_2\} = \emptyset}
            \pr{\ch{v}_i}(h_1, h_2)\;+ & \\
        & + \DP(g_1, g_2)_{i-1,j}\sum_{(h_1, h_2) \in G^2}
            \mathbb{I}_{\{h_1, h_2\} \cap \{g_1, g_2\} \neq \emptyset}
            \pr{\ch{v}_i}(h_1, h_2) & i > 0
    \end{aligned}\right.
\]
Once we have computed the values of this table for $i = |\ch v|$, we
can compute $\pr v$:
\[
    \pr{v}(g_1, g_2) \cong \sum_{j = 0}^{|\ch{v}|} \varepsilon^j \cdot \DP(g_1, g_2)_{|\ch{v}|, j}
\]
Constructing the DP table takes $O\left( \alpha|G|^4 \right)$ time per
block, which dominates the runtime of computing the marginals by this
method. We can further reduce the runtime by directly maintaining the
marginal probability that some single gene appears in each node (in
addition to the probability estimate over pairs of genes $\pr v$):
\[
    S_v(g) = \sum_{g' \in G}\pr{v}(g, g')
\]
Then we can compute the DP as below:
\[
    \DP(g_1, g_2)_{i,j} = \left\{\;\begin{aligned}
        & \mathbb{I}_{j = 0} & i = 0 \\
        & \DP(g_1, g_2)_{i-1,j}\left(
            S_v(g_1) + \mathbb{I}_{g_1 \neq g_2}
            (S_v(g_2) - \pr{\ch{v}_i}(g_1, g_2))
        \right) \;+ \\
        & + \DP(g_1, g_2)_{i-1,j-1}\left(
            1 - (S_v(g_1) + \mathbb{I}_{g_1 \neq g_2}
            (S_v(g_2) - \pr{\ch{v}_i}(g_1, g_2)))
        \right) & i > 0
    \end{aligned}\right.
\]
Computing the DP table in this manner requires only
$O\left( \alpha|G|^2 \right)$ time.

However, as presented, the BP sweep for symbol collection has a memory
complexity of $O\left( |G|^2 \right)$ per block per node, which in
practice is prohibitive even for pedigrees with relatively small
founding populations. To reduce the memory complexity by a factor of
$|G|$, we make the simplifying assumption that
the probability that some vertex $v$ has at least one of a pair of
genes $g_1, g_2$ approximately equals
$S_v\left( g_1 \right) + S_v\left( g_2 \right)$; this permits us to
store only the marginal probabilities over single genes, rather than
the entire distribution over pairs of genes.

The DP values are then calculated as follows:
\[
    \DP(g_1, g_2)_{i,j} = \left\{\;\begin{aligned}
        & \mathbb{I}_{j = 0} & i = 0 \\
        & \DP(g_1, g_2)_{i-1,j}\left(
            S_v(g_1) + \mathbb{I}_{g_1 \neq g_2} S_v(g_2)
        \right) \;+ \\
        & + \DP(g_1, g_2)_{i-1,j-1}\left(
            1 - (S_v(g_1) + \mathbb{I}_{g_1 \neq g_2} S_v(g_2)
        \right) & i > 0
    \end{aligned}\right.
\]

On small pedigrees, this assumption does not produce a decrease in
reconstruction accuracy. We also show that simulations on large
pedigrees, which are impractical with the $O\left( |G|^2 \right)$
per-block memory complexity, perform well.

We also implement a relatively simple parsimony-based symbol
collection step, which greedily takes the genes that entail the fewest
topology errors. 

\section{Simulation Results for BP}\label{sec:bp-sim}

\begin{figure}
    \centering
    \begin{subfigure}{.475\textwidth}
        \centering
        \includegraphics[width=\textwidth]{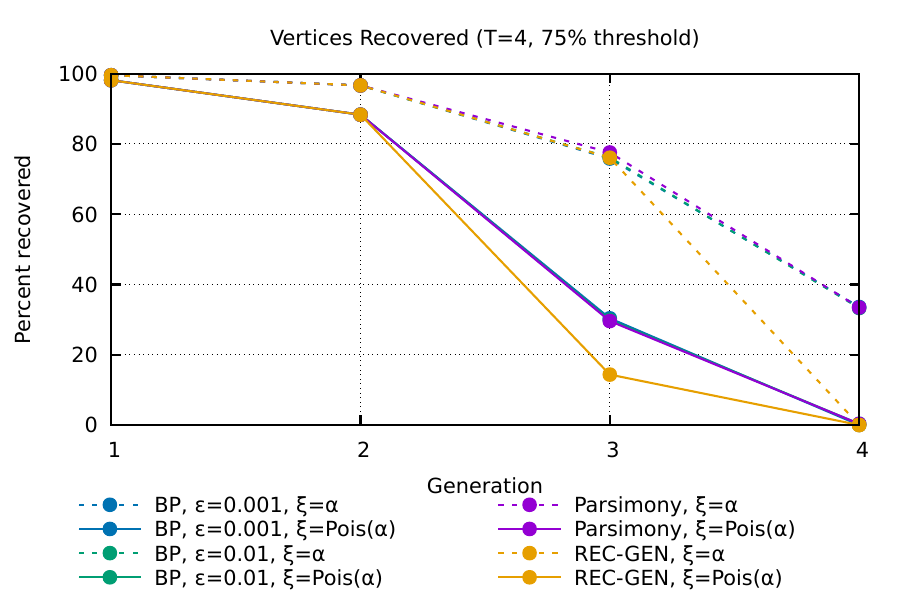}
    \end{subfigure}
    \begin{subfigure}{.475\textwidth}
        \centering
        \includegraphics[width=\textwidth]{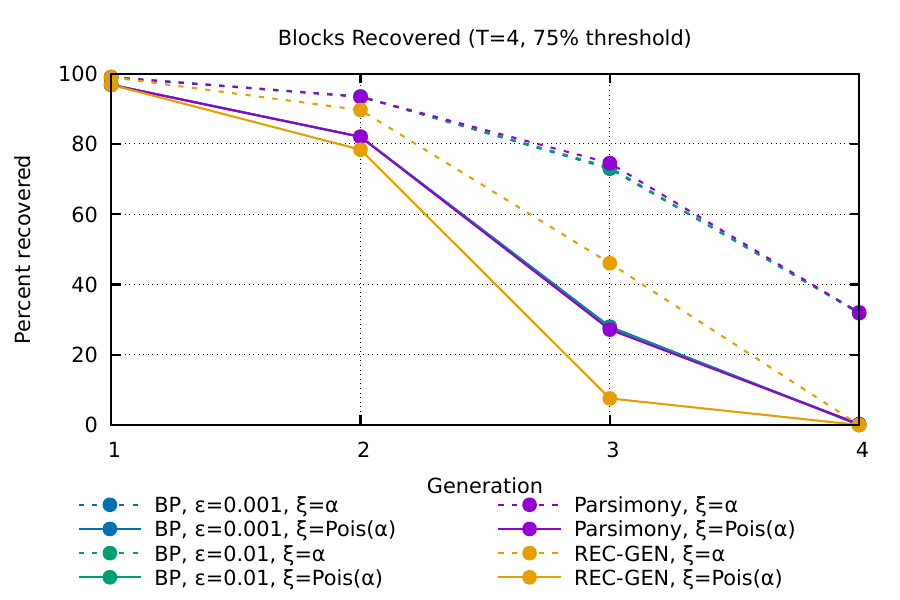}
    \end{subfigure}
    \caption {
        Average percent vertices and blocks 0.75-successfully reconstructed using
        each of four different procedures for symbol collection in
        each generation of $T=4$ pedigrees
    }\label{fig:vert-t4-a75}
\end{figure}

\begin{figure}
    \centering
    \begin{subfigure}{.475\textwidth}
        \centering
        \includegraphics[width=\textwidth]{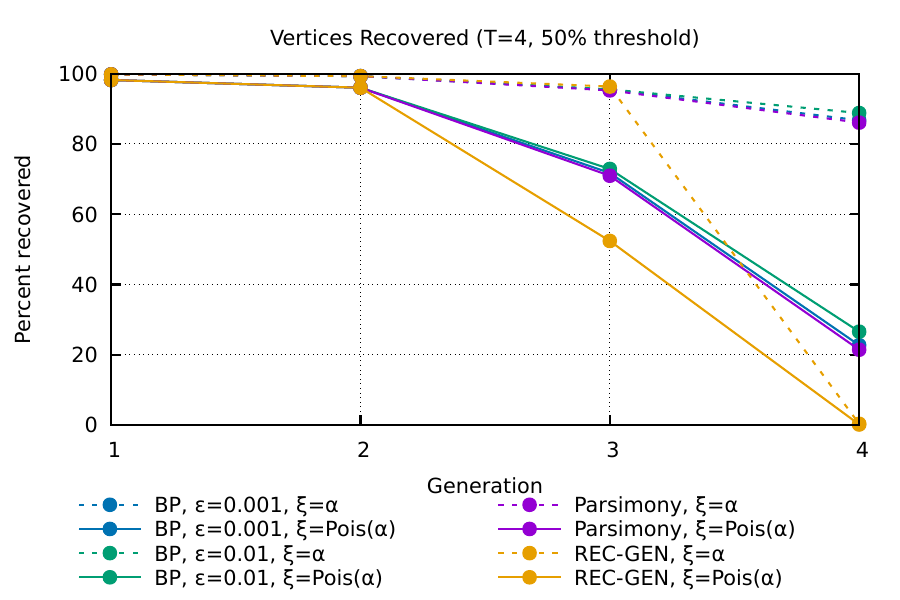}
    \end{subfigure}
    \begin{subfigure}{.475\textwidth}
        \centering
        \includegraphics[width=\textwidth]{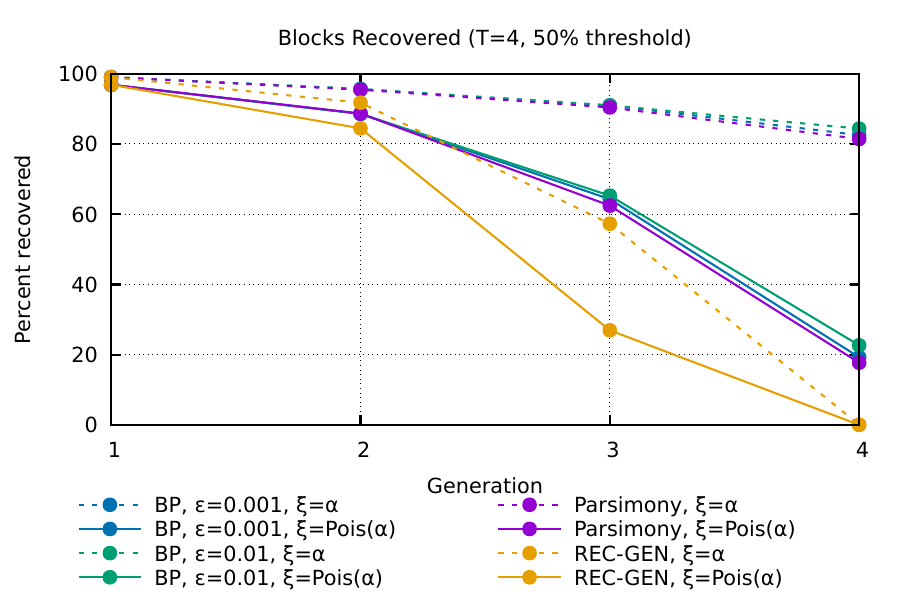}
    \end{subfigure}
    \caption {
        Average percent vertices and blocks 0.50-successfully reconstructed using
        each of four different procedures for symbol collection in
        each generation of $T=4$ pedigrees
    }\label{fig:vert-t4-a50}
\end{figure}

Experiments using simulated data generated as described in
Section~\ref{sec:gen-ped} indicate that using BP or Parsimony instead
of the combinatorial symbol-collection step of \textsc{Rec-Gen}
significantly improves accuracy and permits substantial recovery of
the founding populations of $T=4$ pedigrees without manual
intervention in the siblinghood threshold.
Figures~\ref{fig:vert-t4-a75} and~\ref{fig:vert-t4-a50} show the
reconstruction accuracy of BP with two values of $\epsilon$ (0.01 and
0.001), parsimony, and the original \textsc{Rec-Gen} symbol-collection
step. Parsimony and both instances of BP have similar accuracy, which
past the grandparent generation is significantly better than that of
the original \textsc{Rec-Gen}. BP with $\epsilon=0.01$ tends to
slightly outperform BP with $\epsilon=0.001$ and parsimony. These
results indicate that BP is more robust against inbreeding than the
combinatorial \textsc{Rec-Gen}. While parsimony is a simple
approximation of BP, its reliability decreases when
the distribution of fertilities is non-constant.

\section{Discussion}

The changes to the {\sc Rec-Gen} algorithm of
Ref.~\citen{kim2020efficient} presented in this paper contribute
significant improvements in practical efficiency and accuracy on
simulated pedigree data without sacrificing many of the original
algorithm's theoretical guarantees. We show how to reduce the
complexity of the sibling-identification step from cubic in the size
of the extant population to essentially quadratic while continuing to
use triples as the basis for reconstructing sibling relations and
replace the combinatorial genome reconstruction step with a
significantly faster and more accurate Belief Propagation procedure;
this Belief Propagation procedure is also more accurate than parsimony
when the distribution of fertilities is not constant.

Adaptation of our ideas to real-world data is
beyond the scope of this work as our model assumes well-defined
generations, high fertilities, and no phasing. However, we believe that the
presented contributions can be used in practical tools
for reconstruction.

Source code and simulation data are available at
\url{https://github.com/dvulakh/RecGen}

\section*{Acknowledgments}

This work was partially supported by
Vannevar Bush Faculty Fellowship ONR-N00014-20-1-2826, NSF
award DMS-2031883, MIT UROP, and by a Simons Investigator award
(622132).
\bibliography{paper-draft-ref}

\end{document}